\documentclass[letter]{aa} 

\usepackage{graphicx}
\usepackage{xcolor}
\usepackage{txfonts}
\usepackage{hyperref}
\usepackage[bottom]{footmisc}
\usepackage{natbib}
\usepackage{subfigure}
\usepackage{multirow} 
\renewcommand{\arraystretch}{1.3}
\usepackage{lscape}
\usepackage{tabularx}
\usepackage{float}
\hypersetup{colorlinks=true, allcolors=blue}
\usepackage{orcidlink}
\defcitealias{edenhofer2024}{E24}
\defcitealias{kormann2026}{K26}

\usepackage{placeins}

\begin{document} 

   \title{The superclumps of the local Milky Way}
   \subtitle{Supercloud fragmentation and the sites of star formation}

   \author{
   Lilly A. Kormann\inst{\ref{univie},}\corrauth{lilly.kormann@univie.ac.at} \orcidlink{0009-0009-2174-5363} 
   \and Jo\~ao Alves \inst{\ref{univie}}\orcidlink{0000-0002-4355-0921} 
   \and Emily L. Hunt\inst{\ref{univie}}\orcidlink{0000-0002-5555-8058} 
   }
   \authorrunning{Kormann, L. A., et al.}

   \institute{
    University of Vienna, Department of Astrophysics, T\"urkenschanzstrasse 17, 1180 Wien, Austria\label{univie}
             }

   \date{Received 17 August 2026 / Accepted ...}

  \abstract
  {Using a \textit{Gaia}-based 3D dust map of the solar neighborhood, we analyze the internal structure of the seven local superclouds. We identify quasi-periodic density enhancements along their spines, which we term ``superclumps'' and show that 73\% of the known star-forming regions in the dust map volume can be associated with them. Across the six superclouds with more than one recovered superclump, the spacings are characteristic per cloud: $\sim$150--250\,pc for the Split, Malpolon Cloud and Vela Ridge Cloud, and $\sim$250--380\,pc for the Radcliffe Wave, Natrix Cloud and Sagittarius Spur Extension. The observed separations are two to three times smaller than the $\sim$560\,pc predicted for an isolated self-gravitating cylinder of the same effective diameter, indicating fragmentation under external pressure rather than in isolation. The recurring spacing suggests that giant molecular cloud assembly is not a local, stochastic process, but is instead influenced by the large-scale gravitational fragmentation of the parent superclouds. Regardless of the precise formation mechanism, the superclumps occupy a critical intermediate scale in the hierarchical organization of the interstellar medium, bridging the gap between the kiloparsec-scale gas lanes and the $\sim$10--100\,pc scale of individual giant molecular clouds.}

   \keywords{ISM: structure -- ISM: clouds -- Stars: formation}
   \maketitle
   \nolinenumbers


\section{Introduction}

    The origin of giant molecular clouds (GMCs) remains one of the least constrained problems in star formation theory. Proposed formation mechanisms range from large-scale gravitational instabilities and converging flows in spiral arms to stochastic cloud-cloud collisions and feedback-driven shell fragmentation. Yet, observations have struggled to distinguish between these scenarios, as individual GMCs are typically studied in isolation from their larger-scale environment \citep[see reviews by][]{dobbs2014,chevance2020}. 

    GMCs are the primary sites of star formation in the local Milky Way, yet their accurate spatial distribution has only recently become accessible \citep{zucker2020, alves2020}. The idea that GMCs are embedded in larger structures dates back to \cite{elmegreen1983} and \cite{elmegreen1987}, who identified neutral atomic hydrogen (H\textsc{i}) superclouds in the spiral arms of external galaxies and in the inner Milky Way. They showed that GMCs are not randomly distributed but rather are spatially correlated and embedded within these structures. Despite this connection, superclouds in the immediate solar neighborhood had remained largely unexplored.  

    \citet[hereafter \citetalias{kormann2026}]{kormann2026} used the \textit{Gaia}-based 3D dust map of \cite{edenhofer2024} to close this gap and identified seven superclouds in the solar neighborhood, finding that ${\sim}$90\% of the known star-forming regions \citep[SFRs,][]{zucker2020} and their associated GMCs lie within them. These structures act as a bridge between the smaller-scale GMCs and galactic-scale features. To assess whether there is a correlation between the internal structure of the superclouds and the matched SFRs, we analyze their internal density structure. Using the dataset from \citetalias{kormann2026}, we apply the HOP structure finder \citep{colman2024} with an updated set of parameters to identify high-density regions, match them with known SFRs, and analyze their spatial distribution.

\begin{figure*}[!ht]
   \centering
   \begin{minipage}[c]{0.6\textwidth}
       \centering
       \includegraphics[width=\textwidth]{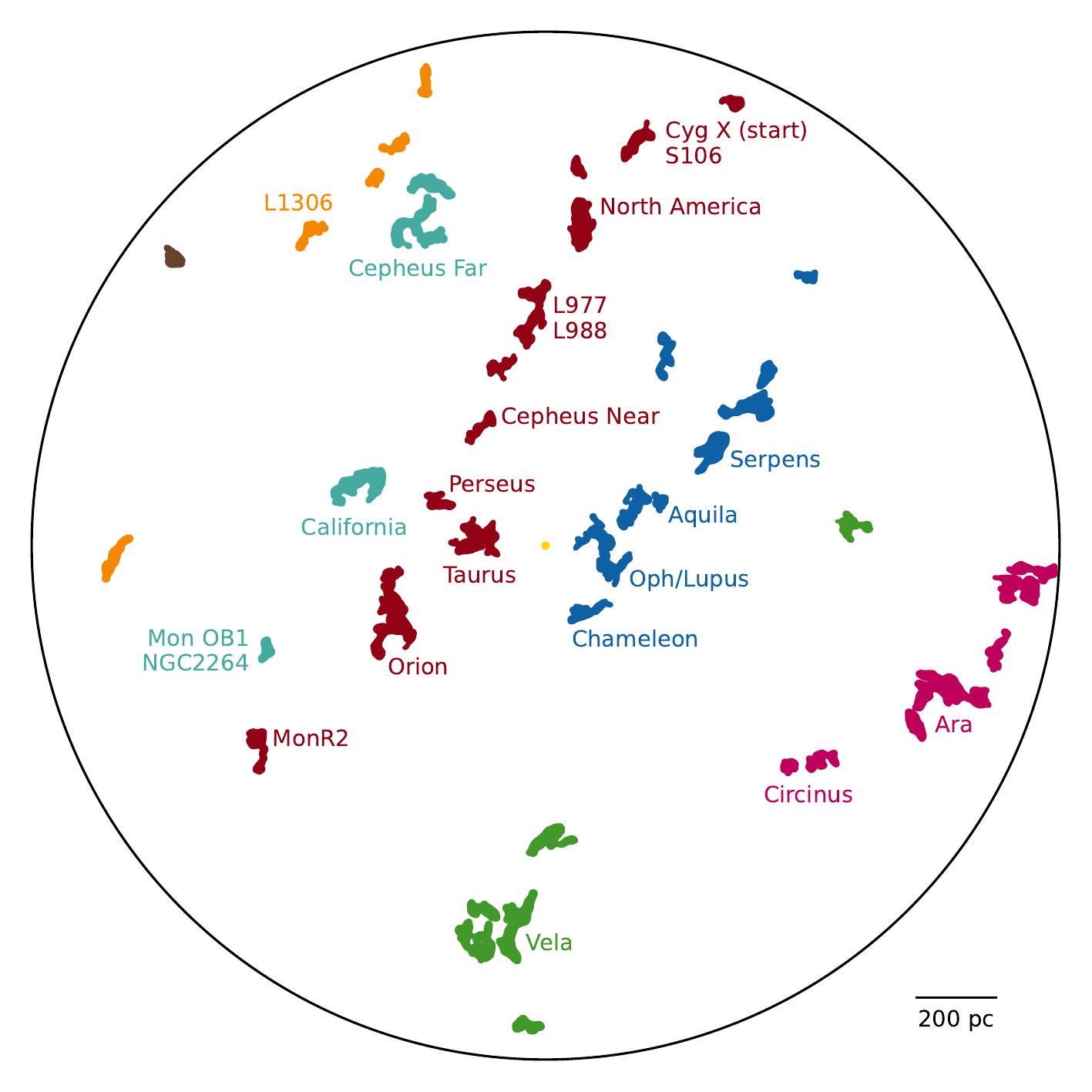}
   \end{minipage}
   \hfill
   \begin{minipage}[c]{0.33\textwidth}
       \centering
       \includegraphics[width=\textwidth]{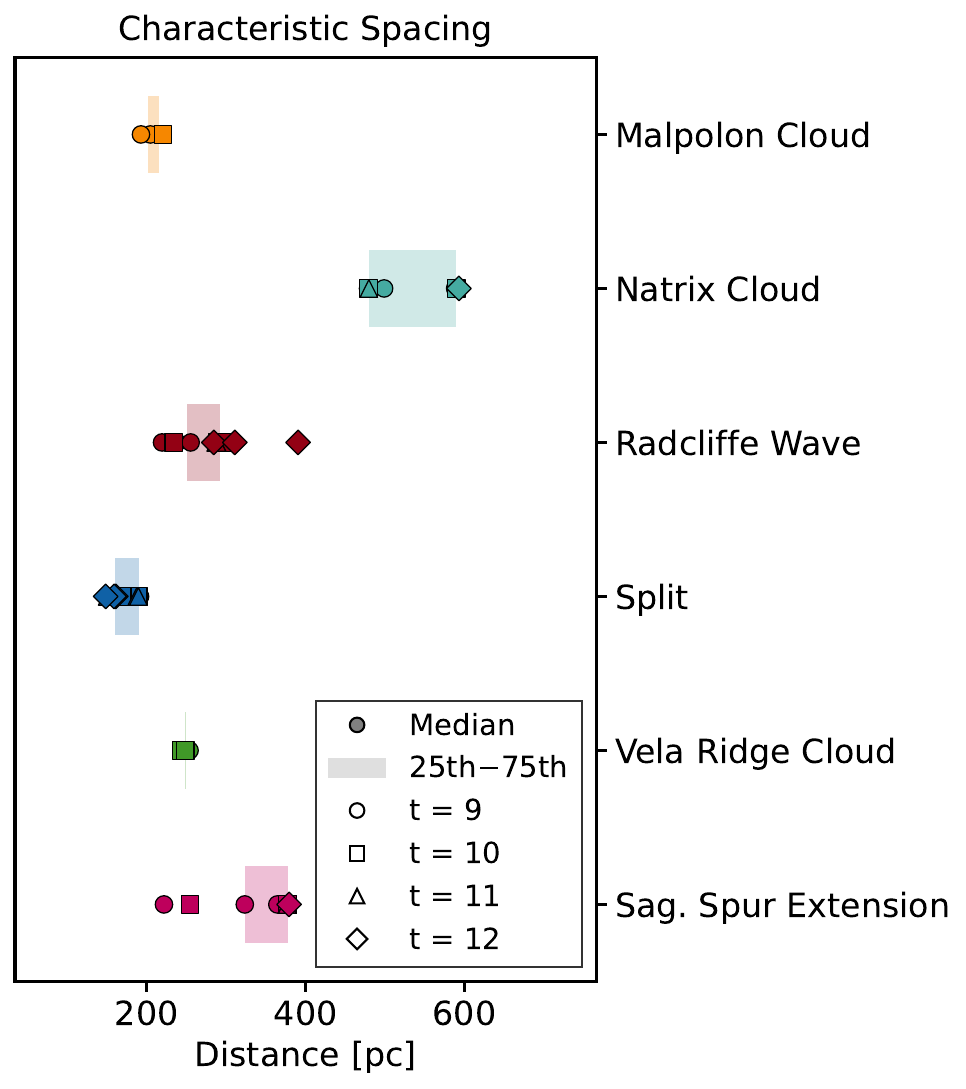}
   \end{minipage}
   \caption{Left: Face-on view of an example HOP segmentation using a minimum density threshold of 9 cm$^{-3}$ and a minimum number of cells of 2000. The colors correspond to the superclouds as defined in \cite{kormann2026}, from left to right: the Anguis Cloud (brown), Malpolon Cloud (orange), Natrix Cloud (turquoise), Radcliffe Wave (red), Vela Ridge Cloud (green), and Sagittarius Spur Extension (pink). The Sun is represented by the yellow dot in the center; the Galactic Center is located to the right. An interactive 3D version of the superclumps is available \href{https://lillykormann.github.io/superclouds_figures/superclumps/t9_n2000_surface.html}{here}. 
   Right: Characteristic superclump spacings for the 12 HOP configurations. Markers indicate different minimum density thresholds, and the shaded area shows the 25th to 75th percentile range. Here, the star-forming region L1335 is not included for the Natrix Cloud, as discussed in Sect. \ref{sec:disc_superclump_spacing}. }
       \label{fig:t9_n2000}
\end{figure*}

\section{Data}
\label{sec:data}

    We used the supercloud dataset from \citetalias{kormann2026}, consisting of the hydrogen volume density distribution derived from the \cite{edenhofer2024} dust map, to analyze their interior structure. HOP identifies structures around local density peaks, excluding voxels below a minimum density threshold, and discarding or merging neighboring structures based on their size and boundary densities. The minimum density threshold (\texttt{t}) was varied from the initial supercloud value of \texttt{t} = 1.3 cm$^{-3}$ up to 12 cm$^{-3}$. The second varied parameter was the minimum number of cells (\texttt{n}), which sets the minimum size of the recovered structures. In Fig. \ref{fig:full_sample_grid} we show the changes when increasing \texttt{t} (vertically) and \texttt{n} (horizontally). The conversion process from unitless extinction to hydrogen volume density and the application of HOP are further detailed in \citetalias{kormann2026}.

    For the analysis, we focused on \texttt{t} = 9 to 12 cm$^{-3}$. 
    We analyzed the full density range from the supercloud detection threshold (1.3~cm$^{-3}$) upward. When reaching the finalized minimum threshold, we find a coherent pattern of denser regions along the spines of the superclouds. 
    The minimum density range ensures that we analyze regions roughly 2 -- 3 times denser than the average supercloud density \citep[$\sim$4 cm$^{-3}$,][]{kormann2026}; lower thresholds recover a more continuous distribution, still following the spines of the superclouds with less distinct fragmentation.
    Each threshold configuration is analyzed using three different \texttt{n}; \texttt{n} = 1000, 1500, and 2000. This traces the denser, larger regions along the spines of the superclouds while filtering out smaller structures that might meet the threshold criteria but are not of interest in this study. 
    The remaining two HOP parameters, peak density (\texttt{p}) and saddle density (\texttt{s}), were set equal to \texttt{t}. While \texttt{p} and \texttt{s} influence the merging behavior and internal structure of the resulting clouds, for the chosen \texttt{t} the recovered clouds remain practically identical. The main differences are the separation of previously merged neighboring clouds, which does not affect the analysis outcome. To simplify the approach, we therefore adopt \texttt{t} = \texttt{p} = \texttt{s}. 

\section{Methods and results}
\label{sec:methods_and_results}

\subsection{Superclump identification}
\label{sec:superclump_identification}

    With the 12 HOP configurations, we obtain a dataset that traces the inner, denser regions of the superclouds across a range of varying parameters. This ensures that selecting a specific configuration does not filter out potential regions of interest. At the same time, the higher parameter combinations visualize the densest regions we can trace using the \cite{edenhofer2024} dust map. An example of the resulting segmentation with \texttt{t}~=~9 ~cm$^{-3}$ and \texttt{n}~=~2000 can be seen in the left panel of Fig.~\ref{fig:t9_n2000}. The superclumps, together with the supercloud contours, are shown in Fig.~\ref{fig:superclouds_superclumps}; the full sample is visualized in Fig.~\ref{fig:full_sample_grid}.

    We label the identified regions along the spines ``superclumps'', following the terminology as described in \cite{williams2000} for the structure of molecular clouds. Many correspond to the location of known SFRs in the solar neighborhood as cataloged in \cite{zucker2020}, who derived homogeneous distances for ${\sim}$60 SFRs within 2.5 kpc from the Sun. This relation between superclouds and SFRs was already mentioned in \citetalias{kormann2026}, with a match rate of ${\sim}$90\% for the full supercloud sample. 
    
    SFRs were matched to superclumps within 50~pc of the superclump surface, a threshold chosen to account for the SFR distance and superclump boundary uncertainties, while remaining consistent across HOP configurations. 
    The additional associations recovered at 50~pc compared to lower tested thresholds lie within the expected spatial distribution of SFRs and do not introduce spurious matches.
    For the superclump volume, we obtained a match rate of 73\%.

    To visualize the superclumps and associated regions, we labeled the SFRs in the left panel of Fig. \ref{fig:t9_n2000}.
    Although the \cite{zucker2020} distance for the Circinus SFR lies beyond the 50 pc threshold, we connected it with the closest superclump in the figure, as distances between 700 and 900 pc are reported for the region \citep[see][and references therein]{kerr2025}, leading to an overlap with the labeled superclump.
    As described in \citetalias{kormann2026}, the Anguis Cloud does not contain any of the known SFRs, which is also shown here.

    Figures \ref{fig:t9_n2000} and \ref{fig:full_sample_grid} reveal that the spatial distribution of superclumps along the spines follows a characteristic spacing, more prominent for the higher \texttt{t} configurations. To quantify this, we measured the superclump spacings as described below. For the Anguis Cloud, we only recovered one superclump and therefore excluded it from the spacing analysis.
    
\subsection{Spacing calculation}
\label{sec:spacing_calculation}

    Several connection conditions were applied to trace the superclump pattern. We grouped clumps that originated from a single structure at lower density thresholds and clumps that were spatially close. As \texttt{t} increases, larger structures progressively fragment. We kept such fragments grouped at higher thresholds to avoid introducing small-scale separations in the analysis.
    Superclumps were then connected sequentially along the spines of their parent supercloud, following the observed face-on orientation \citep[${\sim}30\degr$,][]{kormann2026}. 
    We did not connect superclumps lying parallel to each other perpendicular to the supercloud spine, as seen, e.g., in the Split, where regions appear on both sides of the Aquila Ring \cite[][see Fig. \ref{fig:superclouds_superclumps} at X = 450 pc and Y = 450 pc]{benjamin2024}. 
    In the Malpolon Cloud, superclumps in the upper-left corner sit elevated above the Galactic plane. We excluded connections between these and the lower structures, as external influences (e.g., feedback of high-mass stars as traced by \cite{zari2023} and \cite{pantaleoni2025}) have likely disrupted the pattern.  
    The superclump associated with the Taurus molecular cloud was excluded from the analysis of the Radcliffe Wave, as it likely formed on the opposite side, near the Split, and has traveled to its current position over the last 2 to 6 Myr \citep[Cameren Swiggum, private communication]{zucker2022}.
    
    We calculated the spacings between superclump pairs in the original 3D space as the Euclidean distances between the density-weighted centroids of consecutive superclumps. Spacings were calculated per supercloud only, as we aim to quantify the separation along individual spines.
    
    The spacings were calculated independently for each of the 12 HOP parameter configurations. These configurations are not independent measurements; they serve as a set of thresholds that span a range over which the recovered structures remain physically meaningful, rather than relying on a single configuration.
    For each supercloud and each configuration, we calculated the median of the recovered spacings when at least two spacing values were present, yielding a characteristic spacing range. We calculated the ``core'' range (25th--75th percentile) and the ``full'' range (minimum--maximum). These ranges measure the sensitivity and consistency of the spacings across the parameter choices only and are not confidence intervals, since the 12 configurations all trace the same underlying density field. Due to this and the low number of spacing measurements per supercloud in each configuration, a full superclump spacing analysis is not possible without hand-picking a configuration. 
    
    Before computing the medians, we excluded the largest spacing (${>}800$~pc) in two superclouds: the Malpolon Cloud and the Vela Ridge Cloud. In both cases, spatially isolated superclumps, located at X = 746 pc, Y = 48 pc and X = -1049 pc, Y = -31 pc in Fig. \ref{fig:full_sample_grid} respectively, inflate the characteristic spacing. We set the cutoff to 800~pc, as it lies below the spacings of the isolated superclumps, yet is high enough not to affect the other distance measurements. The large spacings of isolated superclumps are further discussed in Sect.~\ref{sec:disc_superclump_spacing}.

\subsection{Spacing analysis}
\label{sec:spacing_analysis}

    In the right panel of Fig.~\ref{fig:t9_n2000}, we show the characteristic spacing recovered per supercloud, per configuration. A second version can be found in Fig.~\ref{fig:spacing_distribution}. The Split and the Radcliffe Wave recover a spacing range in all 12 configurations, with core ranges of 160--190~pc and 251--292~pc. For the Sagittarius Spur Extension, a spacing range is recovered in 9 of 12 configurations, with a core range of 323--377~pc. Applying the 800~pc cutoff removes the isolated, widely separated superclump distances from the Malpolon Cloud and Vela Ridge Cloud, leaving only 4/12 and 5/12 configurations, but with tight core ranges of 202--215~pc and 248~pc, respectively. Lastly, we recover spacings in 7 out of 12 configurations for the Natrix Cloud, resulting in a core range of 479--589~pc; we further discuss this result, including an SFR without a recovered superclump, in Appendix~\ref{sec:natrix_l1335}. The results for both ranges per supercloud are shown in Table \ref{tab:spacing_table}.

\section{Discussion}
\label{sec:discussion}

\subsection{Supercloud fragmentation}
\label{sec:disc_supercloud_fragmentation}

    Superclouds do not fragment randomly. We find that regions of higher density, termed ``superclumps'', emerge along their central spines and predominantly map to known SFRs in the solar neighborhood. Regarding the \cite{edenhofer2024} dust map volume, roughly 90\% of the known SFRs are located within the superclouds, and 73\% can be associated with the superclumps identified in this work. This correlation points to a link between large-scale, likely gravitational, fragmentation and the sites where stars are born. 
    Quasi-periodic structure of this kind is not new. For example, \citet{henshaw_ubiquitous_2020} found regularly spaced density enhancements throughout the molecular interstellar medium (ISM), coupled to oscillatory gas flows with wavelengths of $0.3-400$ pc, highlighting that the structure of the ISM should not be studied in isolation. 
    What is new here is the parent: The superclumps are substructures within individually mapped hosts, whose length, mass, pitch angle, and mean density are measured, allowing for the spacing to be set against the host’s structural properties for the first time.

    The introduced superclump framework offers a new observational constraint on the origin of GMCs. Their characteristic spacing is consistent with GMC formation not being a purely local, stochastic process but rather influenced by the large-scale fragmentation of the parent supercloud, which seems to pre-select the sites where molecular gas accumulates, and stars can eventually form. 
    This reveals a previously unrecognized hierarchy of interstellar structure in the local Milky Way, with the superclumps representing the progenitor environment of GMCs: A ${\sim}100$~pc concentration of mostly atomic gas in which molecular clouds assemble and ultimately star formation takes place. 

\subsection{Superclump spacing}
\label{sec:disc_superclump_spacing}

    An indicator of this pre-selection for star-formation sites is the characteristic spacing of the superclumps themselves. Applying the 800~pc cutoff (see Sect.~\ref{sec:spacing_calculation}), the superclouds showing the narrowest core ranges are the Vela Ridge Cloud, Split, Radcliffe Wave and Malpolon Cloud, with a core range difference of 10~--~40~pc across the configurations. This is followed by the Sagittarius Spur Extension (core range ${\sim}$50~pc) and the supercloud with the largest core range, the Natrix Cloud, showing roughly 100~pc. The superclumps differ in length and mass, yet we still recover a characteristic spacing for denser regions. This points towards a common fragmentation mechanism operating under local boundary conditions set by the superclouds. 

    The Radcliffe Wave and the Split sit closest to the Sun, which could aid the recovery of superclumps and the resulting characteristic spacing. The lower number of recovered superclumps and therefore measured spacings for superclouds located further from the Sun could reflect different conditions setting the fragmentation scale, or be a result of the decreasing reliability of the underlying data for the \cite{edenhofer2024} dust map, as \textit{Gaia} parallax uncertainties increase with distance. The isolated superclumps excluded by the 800~pc cutoff in the spacing analysis for the Malpolon and Vela Ridge Cloud might therefore partially reflect incompleteness at higher densities in the dust map and distance uncertainties. When excluding these superclumps, both superclouds recover smaller characteristic spacings, sitting at the lower end of the spacing sample. Whether this reflects a physical outlier or an effect of the map incompleteness will be testable with the next generation of dust maps. The Natrix Cloud presents a similar yet distinct case; instead of showing a single isolated superclump, all three recovered clumps are largely separated. Interestingly, we find an additional SFR \citep[L1335,][]{zucker2020} located within the supercloud, yet no superclump was recovered when applying our chosen thresholds. The full analysis, including L1335 as a stand-in superclump, is detailed in Sect. \ref{sec:natrix_l1335}; in short, we recover more than two spacings for each of the 12 configurations, and the core range is shifted to 357--366~pc (full range 356--374~pc), closer to the other supercloud results. 

\subsection{Potential spacing causes}
\label{sec:disc_spacing_causes}

    Regarding scenarios that could explain the observed superclump spacing, the Toomre instability of the galactic disk \citep{toomre1964} can be ruled out on two grounds. It is geometrically incompatible, operating on a full rotating two-dimensional disk and producing fragmentation at kiloparsec scales, whereas the superclumps are chains of density enhancements along the spines of elongated structures. Furthermore, the galactic disk is Toomre-stable in the solar neighborhood \citep[Q $\sim1.5-1.7$,][]{binney2008, george2025}. 
    
    Longitudinal gravitational fragmentation of self-gravitating filaments \citep{ostriker1964, inutsuka1992} is geometrically compatible with the observed pattern. 
    However, the resulting fragmentation scale from the classical, near-critical, isolated-cylinder analysis does not match the superclump scales. 
    The predicted fragmentation corresponds to four times the effective diameter \citep{inutsuka1992},
    we use the geometric mean of the approximate average width (${\sim}180$~pc) and height (${\sim}110$~pc) of the superclouds obtained via principal component analysis (see \cite{kormann2026} for details), yielding ${\sim}140$~pc. This predicted spacing of ${\sim} 560$~pc sits above the observed scales for five of the six superclouds (${\sim}150-380$~pc), and is only close to what we recover for the Natrix Cloud before accounting for L1335. We do not read this near-match for one supercloud as support for the isolated-cylinder prediction.

    This discrepancy is expected, given that superclouds are pressure-confined structures embedded within the multiphase ISM rather than isolated cylinders.
    \citetalias{kormann2026} noted that while the supercloud line masses vary by about a factor of 4, their mean densities vary by only ${\sim} 10\%$, consistent with approximate pressure equilibrium with the surrounding medium. Under these conditions, pressure confinement enables fragmentation in subcritical structures compared to the simpler case of isolated cylinders \citep{fischera2012}. Furthermore, simulations show that for elongated structures larger than ${\sim}100$~pc, the average line mass no longer reliably predicts fragmentation. Instead, fragmentation appears to follow local enhancements of the line mass along the spine, and kiloparsec filaments that are subcritical on average fragment at such local supercritical peaks \citep{pillsworth2025}. 

\section{Conclusion}
\label{sec:conclusion}
    The picture that emerges is one in which superclumps are gravitationally fragmented density enhancements within pressure-confined atomic gas, where subsequent cooling and molecular cloud assembly produce the GMC complexes and SFRs observed today. If this fragmentation picture holds, the regularly spaced SFRs observed along spiral arms in external galaxies \citep{elmegreen1983, elmegreen2018, gusev2022, kostiuk2025} may reflect the same fragmentation process operating under different local conditions, with spacings set by the local gas density and velocity dispersion. Locally, and if the superclumps are gravitational fragments of their parent superclouds, the velocity field along the spine would be expected to oscillate with a wavelength close to the superclump spacing, as shown on smaller scales \citep{henshaw_ubiquitous_2020}. This prediction should be testable with existing CO and H\textsc{i} surveys across the six superclouds with recovered spacings.


\begin{acknowledgements}
Co-funded by the European Union (ERC, ISM-FLOW, 101055318). Views and opinions expressed are, however, those of the author(s) only and do not necessarily reflect those of the European Union or the European Research Council. Neither the European Union nor the granting authority can be held responsible for them. The results in this paper were based on observations obtained with \textit{Gaia}, an ESA science mission with instruments and contributions directly funded by ESA Member States, NASA, and Canada. 
This research has used NASA's Astrophysics Data System, as well as publicly available software libraries, including Astropy \citep{robitaille_astropy_2013}, NumPy \citep{numpy}, matplotlib \citep{matplotlib}, and plotly \citep{plotly}. 
The computational results were obtained using the Austrian Scientific Computing (ASC) infrastructure.
\end{acknowledgements}

\bibliographystyle{aa} 
\bibliography{ref.bib} 

\appendix
\nolinenumbers

\section{Superclouds and superclumps}

\begin{figure}[!h]
    \centering
    \includegraphics[width=\linewidth]{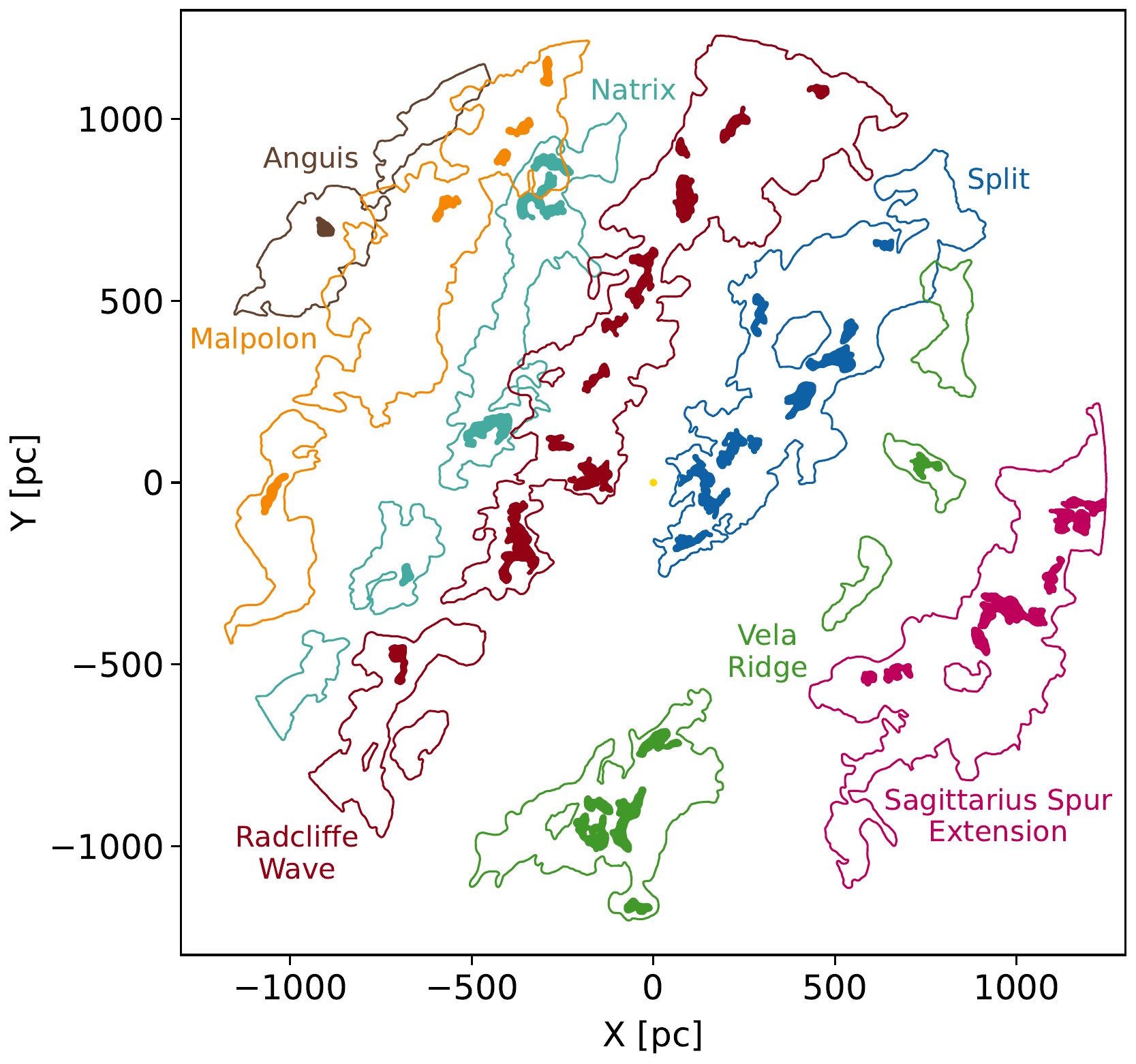}
    \caption{Face-on view of an example HOP segmentation using a minimum density threshold of 9 cm$^{-3}$ and a minimum number of cells of 2000, same as in Fig.~\ref{fig:t9_n2000}. The contour colors correspond to the superclouds as defined in \cite{kormann2026}. The Sun is represented by the yellow dot in the center; the Galactic Center is located to the right.}
    \label{fig:superclouds_superclumps}
\end{figure}

We show the superclumps together with the supercloud contours in Fig.~\ref{fig:superclouds_superclumps}. The figure highlights that superclumps trace chain-like patterns along the supercloud spines, sitting close to their centers rather than the cloud boundaries. The number of recovered superclumps also varies considerably across the superclouds, as discussed in Sect. \ref{sec:disc_superclump_spacing}, from the single superclump recovered for the Anguis Cloud to the extensive coverage for the Radcliffe Wave and the Split.

\section{HOP sample variation}
\label{ap:hop_sample}

\begin{figure*}[ht]
         \centering
         \includegraphics[width=\textwidth]{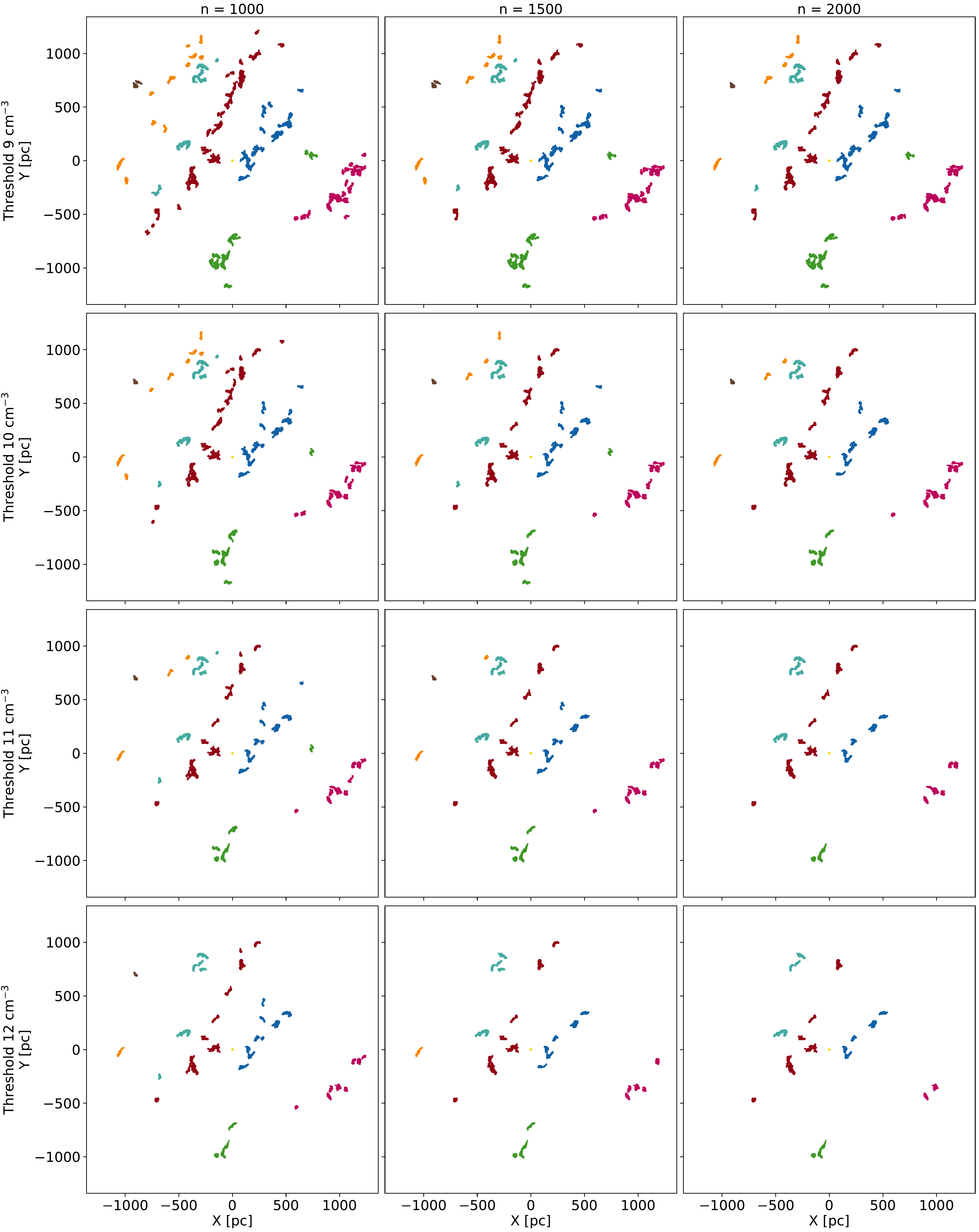}
         \caption{Face-on view of the segmented superclouds with varying density thresholds and minimum cell values per cloud. From top to bottom, a minimum threshold (\texttt{t}) of 9 cm$^{-3}$, 10 cm$^{-3}$, 11 cm$^{-3}$, and 12 cm$^{-3}$. From left to right, the minimum number of cells (\texttt{n}) is 1000, 1500, and 2000 per cloud. The colors correspond to the superclouds as defined in \cite{kormann2026} and shown in Fig.~\ref{fig:superclouds_superclumps}.}
         \label{fig:full_sample_grid}
     \end{figure*}

The full sample of the minimum threshold and minimum number of cell pairs is visualized in Fig. \ref{fig:full_sample_grid}. The variation of parameters highlights the structural changes, with neighboring configurations recovering largely consistent structures. Increasing the thresholds reduces the number of recovered superclumps: only the densest regions remain at the highest minimum threshold of 12 cm$^{-3}$, and only the largest regions remain at a minimum number of cells of 2000.
    
\section{Superclump properties}   
Multiple properties are calculated for the superclumps following \citetalias{kormann2026} and can be found in Table \ref{tab:superclump_properties}.

\begin{table*}[!t]
    \centering
    \caption{Selection of parameters calculated for the superclumps recovered with the \texttt{t} = 9 and \texttt{n} = 2000 configuration.}
    \renewcommand{\arraystretch}{1.23}
    \begin{tabular*}{\textwidth}{@{\extracolsep{\fill}}>{\hspace{4pt}}l l r r r r r r}
        \hline
        ID & Supercloud & Cell Count & Mass [M$_\odot$] & L [pc] & X [pc] & Y [pc] & Z [pc] \\
        \hline
        \hline
        0 & Sagittarius Spur Extension & 18745 & 6.30e+04 & 195.46 & 987.05 & -355.80 & -41.62 \\
        1 & Radcliffe Wave & 17793 & 8.65e+04 & 237.82 & -371.30 & -167.68 & -110.79 \\
        2 & Vela Ridge Cloud & 13990 & 4.71e+04 & 179.84 & -74.10 & -931.65 & 12.22 \\
        3 & Radcliffe Wave & 12291 & 4.26e+04 & 130.89 & 86.58 & 787.01 & -7.72 \\
        4 & Split & 11737 & 3.99e+04 & 139.56 & 501.29 & 336.67 & 4.16 \\
        5 & Split & 11695 & 4.67e+04 & 124.81 & 408.93 & 234.40 & 31.95 \\
        6 & Split & 11311 & 6.45e+04 & 193.70 & 146.63 & -15.73 & 31.33 \\
        7 & Natrix Cloud & 10917 & 4.02e+04 & 128.42 & -304.48 & 774.63 & 27.75 \\
        8 & Radcliffe Wave & 9427 & 4.20e+04 & 145.71 & -168.39 & 12.59 & -39.87 \\
        9 & Natrix Cloud & 8509 & 3.11e+04 & 144.78 & -443.66 & 151.15 & -74.88 \\
        10 & Natrix Cloud & 7468 & 3.33e+04 & 146.60 & -276.57 & 875.38 & 82.97 \\
        11 & Radcliffe Wave & 7081 & 2.27e+04 & 174.69 & -28.79 & 567.41 & 40.03 \\
        12 & Split & 6718 & 2.56e+04 & 115.54 & 217.03 & 95.41 & 15.70 \\
        13 & Sagittarius Spur Extension & 4979 & 1.71e+04 & 122.46 & 1190.88 & -61.06 & -29.49 \\
        14 & Radcliffe Wave & 4939 & 1.83e+04 & 107.02 & -702.07 & -482.92 & -190.52 \\
        15 & Sagittarius Spur Extension & 4860 & 1.86e+04 & 85.86 & 900.42 & -435.99 & -23.24 \\
        16 & Vela Ridge Cloud & 4781 & 1.68e+04 & 104.64 & -147.14 & -980.50 & 13.71 \\
        17 & Sagittarius Spur Extension & 4499 & 1.42e+04 & 107.44 & 1095.94 & -265.98 & -7.23 \\
        18 & Sagittarius Spur Extension & 4306 & 1.38e+04 & 68.05 & 1126.87 & -106.94 & -21.88 \\
        19 & Radcliffe Wave & 4040 & 1.63e+04 & 113.79 & -155.90 & 288.49 & 126.51 \\
        20 & Radcliffe Wave & 4007 & 1.34e+04 & 112.45 & 221.92 & 983.59 & 11.54 \\
        21 & Malpolon Cloud & 3914 & 1.23e+04 & 91.84 & -573.58 & 762.37 & -10.09 \\
        22 & Radcliffe Wave & 3629 & 1.93e+04 & 93.21 & -261.74 & 105.10 & -99.44 \\
        23 & Sagittarius Spur Extension & 3565 & 1.21e+04 & 68.70 & 1179.23 & -107.51 & 39.17 \\
        24 & Anguis Cloud & 3435 & 1.08e+04 & 76.88 & -903.01 & 700.53 & 33.75 \\
        25 & Split & 3431 & 1.27e+04 & 114.01 & 291.98 & 463.13 & 23.48 \\
        26 & Malpolon Cloud & 3102 & 1.16e+04 & 126.27 & -1050.89 & -31.68 & 82.30 \\
        27 & Vela Ridge Cloud & 3047 & 9.89e+03 & 84.92 & 747.49 & 47.45 & 85.27 \\
        28 & Vela Ridge Cloud & 2927 & 9.88e+03 & 82.97 & -150.84 & -888.22 & 8.44 \\
        29 & Vela Ridge Cloud & 2813 & 9.63e+03 & 78.05 & 26.31 & -709.87 & 49.72 \\
        30 & Malpolon Cloud & 2774 & 9.94e+03 & 57.72 & -414.12 & 895.27 & 29.56 \\
        31 & Vela Ridge Cloud & 2744 & 8.16e+03 & 71.70 & -43.32 & -1168.79 & 87.33 \\
        32 & Vela Ridge Cloud & 2730 & 1.05e+04 & 106.37 & -3.36 & -713.02 & -6.10 \\
        33 & Split & 2684 & 1.48e+04 & 101.20 & 95.52 & -158.43 & -57.93 \\
        34 & Sagittarius Spur Extension & 2541 & 1.17e+04 & 61.81 & 595.61 & -537.74 & -50.61 \\
        35 & Split & 2518 & 6.84e+03 & 74.93 & 538.22 & 418.17 & 2.48 \\
        36 & Radcliffe Wave & 2507 & 8.59e+03 & 57.76 & 77.70 & 918.75 & -22.74 \\
        37 & Malpolon Cloud & 2384 & 7.11e+03 & 83.49 & -291.97 & 1132.54 & 210.41 \\
        38 & Vela Ridge Cloud & 2364 & 7.04e+03 & 85.68 & -188.74 & -950.76 & 16.73 \\
        39 & Malpolon Cloud & 2332 & 6.88e+03 & 76.22 & -359.53 & 977.66 & 238.57 \\
        40 & Split & 2233 & 9.21e+03 & 76.38 & 278.88 & 107.03 & -20.58 \\
        41 & Split & 2223 & 1.15e+04 & 119.87 & 104.32 & -162.81 & 2.30 \\
        42 & Radcliffe Wave & 2155 & 6.13e+03 & 55.88 & 461.25 & 1076.47 & -60.15 \\
        43 & Natrix Cloud & 2138 & 8.95e+03 & 61.88 & -679.47 & -255.79 & 15.26 \\
        44 & Split & 2104 & 7.71e+03 & 60.20 & 637.14 & 654.48 & -67.16 \\
        45 & Sagittarius Spur Extension & 2100 & 6.69e+03 & 80.41 & 668.30 & -523.96 & -14.20 \\
        46 & Radcliffe Wave & 2054 & 7.72e+03 & 76.94 & -105.31 & 438.14 & 25.48 \\
        \hline
    \end{tabular*}
    \label{tab:superclump_properties}
    \tablefoot{Each superclump is assigned a superclump ID and has listed which supercloud it is part of. The column cell count indicates the number of voxels per structure; the mass is calculated using all superclump voxels; L denotes the cloud length calculated with PCA. The columns X, Y, and Z are the unweighted mean coordinates per cloud. The characteristic size of a superclump is $100\pm40$ pc. }
\end{table*}

\section{Spacing analysis}

\begin{figure}[hbt!]
        \centering
        \includegraphics[width=\linewidth]{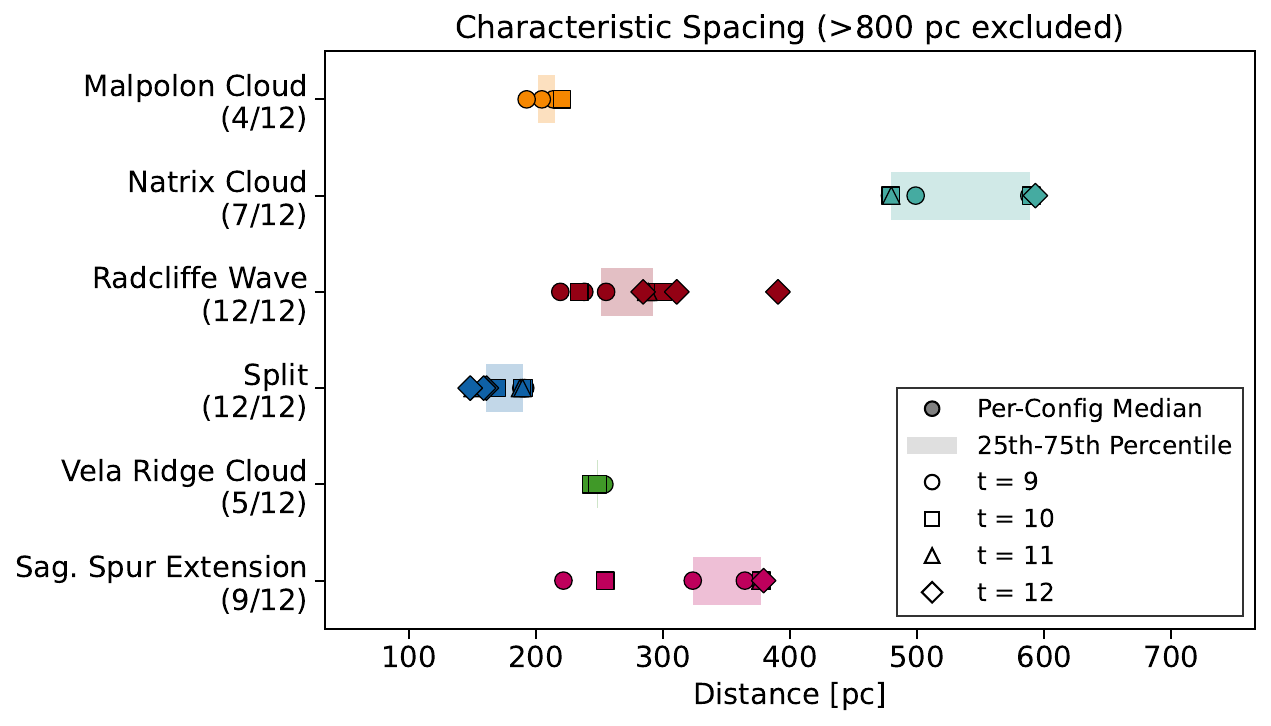}
        \caption{Characteristic superclump spacings recovered per supercloud. Points show the median spacings for each of the 12 HOP configurations; the shaded area shows the 25th to 75th percentile range; the different markers correspond to different minimum density thresholds. Superclouds are labeled with the number of configurations that recovered at least two spacing values among all configurations. This is a descriptive measure of consistency across HOP configurations, not a confidence interval.}
        \label{fig:spacing_distribution}
    \end{figure}

In the right panel of Fig.~\ref{fig:t9_n2000}, we show the characteristic spacing recovered per supercloud; a second version can be found in Fig.~\ref{fig:spacing_distribution}. Each row shows the per-configuration medians and the 25th--75th percentile ``core'' range across the 12 configurations. This is a purely descriptive summary of the recovered medians. As the 12 configurations are not fully independent, we do not treat the resulting range as a confidence interval. Superclouds are labeled with the number of configurations that recover at least two spacing values from the total sample of 12. Table~\ref{tab:spacing_table} shows the ``core'' and ``full'' range results obtained for each supercloud. For the Radcliffe Wave and Sagittarius Spur Extension, the wider spread reflects changes introduced by the different thresholds, shifting the medians to both higher and lower values. The wide separation for the Natrix Cloud is discussed in Appendix~\ref{sec:natrix_l1335}.

\section{The case of L1335}
\label{sec:natrix_l1335}
For the spacing analysis, we focus on calculating distances between the identified superclumps. For each parameter configuration, we recover roughly the same set of superclumps for the Natrix Cloud, each with an associated SFR. A peculiarity regarding Natrix is the SFR L1335 \citep{zucker2020}, which is located within the supercloud, yet we recover no superclump when applying our chosen thresholds. With just the superclump sample, we recover the largest characteristic spacing for the Natrix Cloud out of the six eligible superclouds, with a core range of 479--589~pc in 7 out of the 12 configurations (see Table \ref{tab:spacing_table}). 

We rerun the spacing analysis, including the coordinates of L1335 as a stand-in for a potential superclump. Reasons for it not being recovered could be that it has already been dispersed by stellar feedback, or because HOP misses it due to dust map incompleteness at higher densities and greater distances from the Sun. 

Including L1335 leads to each configuration recovering more than two spacings, and shifts the characteristic spacing down to 357--366~pc (full range 356--374~pc). This happens as L1335 breaks the gap between the upper two superclumps, located at X = -291~pc, Y = 825~pc and X = -447~pc, Y = 148~pc, shown in the top right panel of Fig. \ref{fig:full_sample_grid}, with L1335 sitting roughly halfway between the two at X = -405~pc, Y = 502~pc.

    \begin{table}[hbt!]
        \centering
        \caption{Characteristic spacing recovered per supercloud when applying the 800~pc cutoff. }
        \label{tab:spacing_table}
        \begin{tabular*}{0.5\textwidth}{@{\extracolsep{\fill}}>{\hspace{4pt}}lcc}
        \hline
        Supercloud & Core Range [pc] & Full Range [pc] \\
        \hline \hline
        Split                       & 160--190 & 148--192 \\
        Radcliffe Wave              & 251--292 & 219--390 \\
        Sag. Spur Extension         & 323--377 & 222--379 \\
        Vela Ridge Cloud            & 248--248 & 244--254 \\
        Malpolon Cloud              & 202--215 & 193--221 \\
        Natrix Cloud                & 479--589 & 479--593 \\
        Natrix Cloud (+ L1335)      & 357--366 & 356--374 \\
        \hline
        \end{tabular*}
        \tablefoot{The ``core'' range describes the 25th--75th percentile range, the ``full'' range shows the minimum to maximum range. For the Natrix Cloud, we show the spacings for the superclumps only, and the spacing when the star-forming region L1335 is included, as discussed in Appendix \ref{sec:natrix_l1335}. We adopt the L1335 inclusive range as the representative range for the Natrix Cloud. }
    \end{table}

    \begin{figure}
        \centering
        \includegraphics[width=\linewidth]{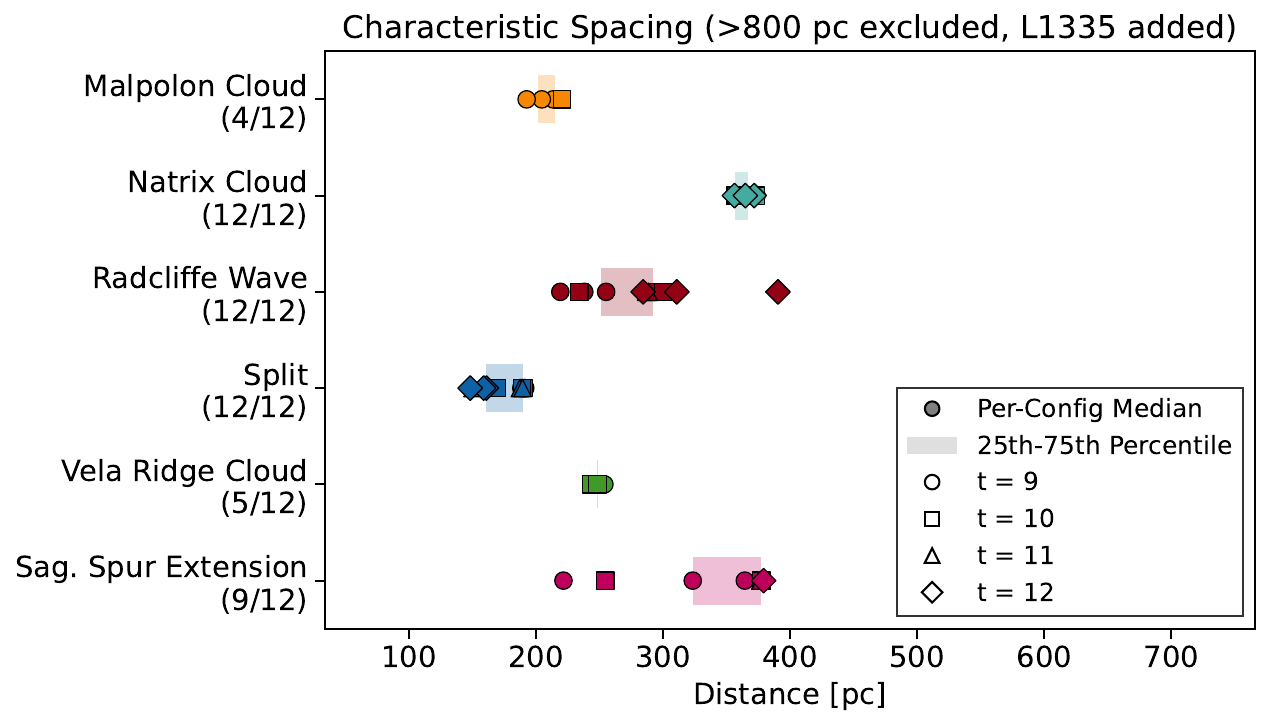}
        \caption{Same as Fig.~\ref{fig:spacing_distribution}. Here we include L1335 as a stand-in superclump for the Natrix Cloud, decreasing the characteristic spacing from 479--589~pc (7/12 configurations) to 357--366~pc (12/12 configurations).}
        \label{fig:spacing_distribution_natrixl1335}
    \end{figure}

\end{document}